\documentstyle[12pt]{article}
\begin{document}
\hspace{11cm}{\bf PUTP-94-08}

\vspace{0.20in}

\begin{center}
\vspace{0.20in} {\Large {\bf Possible Effects of Color Screening
and Large String Tension in Heavy Quarkonium Spectra }}\\

\vspace{0.4in} {\bf Yi-Bing Ding$^{1,2}$ \ \ Kuang-Ta Chao$^{1,3}$ \
\
Dan-Hua Qin$^3$}\\ \vspace{0.3in}

1.{\sl CCAST (World Laboratory), Beijing 100080, China}\\

2.{\sl Department of Physics, Graduate School,}

{\sl \ Academia Sinica, Beijing 100039, China}\\

3.{\sl \ Department of Physics, Peking University, Beijing 100871, China}\\

\vspace{2cm}

{\bf Abstract}
\end{center}

{\small \ Possible effects of the color screened
confinement potential are investigated. Color screened linear
potential with a large string tension
$T=(0.26-0.32)GeV^2$
is suggested by a study of the $c\bar c$ and
$b\bar b$
spectra. The $\psi (4160)$ and $\psi (4415)$ are respectively assigned
as
the $\psi (4S)$-dominated and the $\psi (5S)$ $c\bar c$ states.
Satisfactory results for the  masses and
leptonic widths (with QCD radiative corrections) of $c\bar c$ and
$b\bar b$ states are
obtained.}

\newpage

The string tension is a most fundamental physical quantity in quark
confinement. It is argued based on a rotating string picture that the
 string
tension $T$ is related to the Regge slope $\alpha ^{\prime }$ by $^{[
1]}$

$$
T=\frac 1{2\pi \alpha ^{^{\prime }}}=0.18GeV^2,\eqno(1 )
$$
with the experimental value for $\alpha ^{^{\prime }}=0.9(GeV)^{-2}
$. This
result is supported phenomenologically by the heavy quarkonium
spectrum when
identifying the string tension with the slope of the linear
confinement
potential between a heavy quark-antiquark ($Q\overline{Q}$)
pair$^{[2]}$.
Lattice QCD calculations for the string tension is not conclusive,
because
one needs to estimate the lattice scale $\Lambda _L$ in physical
units, even
if one has obtained from lattice calculations the value for $c$ which
 is
related to $T$ via

$$
c=\frac{\Lambda _L}{\sqrt{T}}.\eqno(2 )
$$
In ref.$[3]$ the string tension is estimated (though with some
theoretical
uncertainty) to be
$$
T=(0.33_{-0.23}^{+0.82})GeV^2\eqno(3 )
$$
with typical values
$$
c=(7.5\pm 0.5)\times 10^{-2},\qquad \Lambda _{\overline{MS}%
}=(200_{-80}^{+150})MeV,\eqno( 4)
$$
although this estimate has some uncertainties, e.g., the value of $c$ is
calculated in the quenched approximation while the experimental value
of $\Lambda_{\overline{MS}}$ is obtained with dynamical fermions (the
more recent data give $\Lambda_{\overline{MS}}^{(5)}=(195+65-50)MeV$
$^{[4]}$). The central value in Eq.(3) is significantly larger than that
given in
Eq.(1), but they may be still consistent with each other when the
large
error involved is reduced.

On the other hand, while the potential model, which assumes a linear
confinement potential plus a one-gluon exchange potential, is generally
successful for $c\bar c$ and $b\bar b$ spectroscopy, some
problems
may remain. One of them is concerned with the assignment of $\psi
(4160)$
and $\psi (4415).$ With the linear confinement potential, $\psi
(4160)$ and
$\psi (4415)$ are usually assigned as the $2D$ and $4S$ states
respectively.
However, experimentally the $\psi (4160)$ has a quite large leptonic
width
$^{[4]}$ $\Gamma _{ee}=0.77\pm 0.23keV$, comparable to that for the
$3S$
state $\psi (4040)$, but in the nonrelativistic limit the $2D$ state
will be
forbidden to decay to $e^{+}e^{-}$. Neither the $S-D$ mixing
nor the coupled
channel models can consistently solve this problem$^{[5]}$. Moreover,
for $\psi (4415)$, with the linear potential the leptonic width
of $4S$ state is usually predicted to be larger by more than a factor
of two than the observed value of $\psi (4415)$, hence the assignment
of $\psi (4415)$ as $4S$ state is also problematic.
The way to solve this problem is probably to assign the $\psi (4160)$
 and
$\psi (4415)$ as the $4S$ and $5S$ states respectively, but this must
require
the linear potential to be softened at large distances.
In addition, with the linearly rising potential the calculated masses and
leptonic widths for highly
excited $b\bar b$ states (e.g. the $6S$ state) usually are also
larger than their observed values.
It is clear that for a pure linear
potential all the wave functions (S-wave) at origion take the same value.
Therefore, if the $Q\bar Q$ potential keeps
linearly rising at large distances the leptonic widths
for highly excited states will gradually approach to a constant value
when the linear potential becomes dominant over
the short ranged Coulomb potential. However, the experimental data
for both $c\bar c$ and $b\bar b$ do not show this tendency at all.
On the contrary, as argued above the data stronly indicate a softening
effect on the linear potential.

Indeed, it is expected (e.g. in the color flux tube picture for
confinement) that at large distances the creation of a
light quark
pair will screen the static color sources of $Q\overline{Q}$, and
will
therefore flatten the linear potential$^{[6]}$. Some recent lattice QCD
calculations
with dynamical fermions seem to indicate that the color screening
effects on
the linear potential do exist at large distances$^{[7]}$.

By taking into account the color screening effect, the $Q\overline{Q}$
potential may be modified and take the form
$$
V(r)=-\frac{4\alpha _s}{3r}+Tr(\frac{1-e^{-\mu r}}{\mu r} ),\eqno(5)
$$
where the first term on the right hand side is the usual one gluon
exchange
Coulomb potential, and the second term is the screened confining
 potential with
a screening parameter $\mu .$ The potential will keep linearly
rising up to
a distance $r\leq \mu ^{-1}$ ( $r\leq (1-2)fm$ for $\mu =(0.2-0.1)
GeV$ ) and
then gradually become flattened and eventually reach a constant value
 $\frac T\mu $ .

Since the presence of the Cornell potential$^{[8]}$, many
improved potentials have been suggested. In particular, the running
coupling constant $\alpha_{s}$ (see e.g. ref.[9]) and one loop QCD
radiative corrections $^{[10]}$ have been successfully incorporated
into the $Q\bar Q$ potential. On the other hand, however,
the understanding
of the confining potential is still poor, though some phenomenological
confining potentials have been considered to improve the fit to the
heavy quarkonium spectra including that of the higher lying
states$^{[11,12]}$.
Since color confinement is the most important part of dynamics in hadron
physics, it is necessary to have further studies regarding
the confining potential together with the $Q\bar Q$
spectroscopy.
In this connection the screened confining potential expressed in (5),
i.e., $V_{sc}(r)=Tr(\frac{1-e^{-\mu r}}{\mu r} )$
may differ from many other
phenomenological potentials (e.g. various power law potentials, see refs.
[11,12] and references therein) in the following respects:

(i) It is indicated by some lattice QCD calculations with dynamical
fermions (see, e.g., ref.[7]).

(ii) It keeps linearly rising up to about one $fm$ and then gradualy
becomes a constant at large distances, and therefore it incorporates
the large distance asymptotic behavior of color screening
into the linear confinement in a natural manner. This is well motivated
theoretically.

(iii) As noted previously$^{[13]}$, the inclusion of color screening is
connected to the removal of the infrared divergences of the $Q\bar Q$
interaction kernel in the momentum space. In fact, in momentum space the
screened confining potential reads$^{[13]}$
$$
V_{sc}( \stackrel{\rightharpoonup }{p})=-\frac T \mu
\delta ^3( \stackrel{\rightharpoonup }{p})+\frac T {\pi
^2}\frac 1{( \stackrel{\rightharpoonup }{p}^2+\mu ^2)
^2}.\eqno (6)
$$
In view of the regularization of the linear potential in momentum space,
the form of (6) and hence (5) seem to be quite natural and unique.
The non-vanishing
value of the cut off $\mu$ is expected to be related to the
polarization of dynamical light quark pairs.

Although the exact form of confinement interaction has not been
analytically derived from the first principles of QCD, we believe
that the screened confining potential $V_{sc}$ expressed in (5)
should be a better candidate for describing confinement than many
other potentials.
Potential (5) may have phenomenological implications and it has been
used in the study of heavy flavor mesons.$^{[13]}$. It will be
interesting to have further phenomenological investications regarding
this potential.

In the following we will use potential (5) to calculate the
$c\overline{c}$
and $b\overline{b}$ mass spectra, and then find the possible
phenomenological values for the string tension $T.$ As the first trial
in a
previous paper$^{[14]},$ we used
$$
T=0.21GeV^2,\quad \alpha _s=0.51,\quad \mu =0.11GeV,\quad m_c=
1.4GeV,\eqno(7 )
$$
as inputted parameters to solve the nonrelativistic Schr\"{o}dinger
equation with
potential (5). The obtained mass spectrum is satisfactory with $\psi
(4160)$
and $\psi (4415)$ assigned as $\psi (4S)$ and $\psi (5S)$ respectively.
However, there are two problems for (7). The first one is that with the
same
value for $T$ and $\mu $ and a smaller value for $\alpha _s$ , we
cannot
find good result for the $b\bar b$ mass spectrum. The second
is
that the value of $\alpha _s$ in (7) seems too large, not compatible
with the present value of QCD scale parameter (see below).

We now find that with
$$
T=0.32GeV^2,\quad \alpha _s=0.306,\quad \mu =0.156GeV,\quad m_c
=1.6GeV \eqno( 8)
$$
for $c\overline{c},$ and
$$
T=0.32GeV^2,\quad \alpha _s=0.275,\quad \mu =0.132GeV,\quad m_b
=4.8GeV \eqno(9)
$$
for $b\overline{b}$, as the the parameters in potential (5) to solve
the
Schr\"{o}dinger equation, good results for both $c\overline{c}$ and
$b\overline{b}$ can be obtained. The calculated masses and leptonic
widths for $c\overline{c}$ states are shown in Table 1, and for
$b\overline{b}$
states in
Table 2. The experimental data are given by the Particle Data
Group$^{[4]}$.
The leptonic widths are calculated using the nonrelativistic
 expressions
without QCD radiative corrections ( $\Gamma _{ee}^0$) and with QCD\
radiative corrections ( $\Gamma _{ee}$)%
(see, e.g., refs.[8,15,16])
$$
\Gamma _{ee}^0=16\pi \alpha ^2e_Q^2\frac{\left| \Psi (0)\right| ^2}
{M^2},%
\eqno(10)
$$
$$
\Gamma _{ee}=\Gamma _{ee}^0(1-\frac{16}{3\pi }\alpha _s(m_Q)),\eqno(11 )
$$
where $\alpha_s(m_Q)$ stands for the coupling constant at the $Q\bar Q$
mass scale, and it can be determined in the time-like processes of
heavy quarkonium decays. Here we use
$\alpha _s(m_c)=0.28$ for $c\overline{c}$ and $\alpha _s(m_b)=0.19$
for $b
\overline{b}^{[16]}.$ These values of the running coupling constant
are
consistent with the QCD scale parameter $\Lambda _{\overline{MS}}\approx
200MeV^{[16]}$.

{}From Table 1 we can see that the $\psi (4160)$ and $\psi (4415)$ are
assigned as the $4S$ and $5S$ states. The predicted leptonic widths
for these two
states are in excellent agreement with data, whereas in the usual
potential
models without color screening effects the $\psi (4160)$ and
$\psi (4415)$
are assigned as $2D$ and $4S$ states and then $\psi (4160)$ would have
zero
leptonic width ( in the nonrelativistic limit ) and $\psi (4415)$
would have
a leptonic width of, say $1.1keV^{[8]}$, too large by more than a
factor of
two than the observed value $(0.47\pm 0.10)keV.$ As for the mass of
the $4S$ state, the predicted value is higher than $\psi (4160)$ by
$100MeV$, and this could be due to the neglect of $S-D$ mixing and
coupled channel effects. In any case, if $\psi (4415)$ is the
$5S$ state, the $\psi (4160)$ must be a $4S$-dominated state with
possibly some mixed components of $2D$ and virtual charmed meson pairs. It
might be interesting to note that in these assignments the $c\bar c$
would have an anomalous mass relation that $m(4S)-m(3S)$ is smaller than
$m(5S)-m(4S)$. Exactly the same anomalous mass relation is also
observed for the $b\bar b$ states$^{[4]}$.
These anomalous mass relations may imply that in the energy region
just above thresholds of many opened channels (e.g., in $3.8-4.3 GeV$
for $c\bar c$) the masses of
resonances can be significantly distorted.
Of course, in explaining these difficulties there could be other
possibilities such as the $c\bar{c}q\bar{q}$ states$^{[17]}$ or
$c\bar{c}g$ states$^{[18]}$. However, these states in general do
not seem to have large enough leptonic widths to be the $\psi (4160)$,
because their couplings to the photon are expected to be suppressed.
In Table 1 the
predicted
mass for $\psi (3S)$ is now $4.03GeV$, much closer to its observed
value
than $4.11GeV^{[8]}$ predicted by usual potential models. Moreover,
in Table
1 the predicted leptonic widths for $\psi (1S)$ and $\psi (2S)$ also
agree
with data.

{}From Table 2 we see that in general the calculated masses and
leptonic
widths for the $b\overline{b}$ states are also in good agreement with
data.
As a result, using potential (5) with parameters (8) and (9), the
obtained $c
\overline{c}$ and $b\overline{b}$ spectra are remarkably improved.

We have also tried to fit the spin-averaged mass spectrum using
potential (5).
Here the
spin-averaged masses for the $S$ wave states mean the masses before
hyperfine splittings, e.g., for $c\bar c$
$m(1S)=\frac 1{4}[3m(J/\psi)+m(\eta_c)]$. For other $S$ wave states,
because of the lack of observed values for the $0^{-}$ mesons, we use
calculated hyperfine splittings (see (12)) and observed $1^{-}$ meson
masses to determine the spin-averaged masses.
We find that with slightly adjusted parameters (e.g., a slightly
smaller string tension and a slightly larger $\alpha_s$)
we can get good fit for the spin-averaged $c\bar c$
mass spectrum and leptonic widths. Again, the assignments of
$\psi (4160)$
and $\psi (4415)$ as the $4S$ and $5S$ $c\bar c$ states seem to require
a screened confining potential with a large string tension.

We have also used a modified Coulomb potential (with a running
coupling constant $\alpha_{s}(r)$) and the screened confining
potential to fit the heavy quarkonium spectra, and the obtained results
are similar to that obtained with the fixed coupling constant
$\alpha_{s}$. Namely, a large string tension with color screening
is still needed if a good fit for the higher excited
states is required.
In another words, taking a running $\alpha_{s}$ does not change the
basic feature of our observation on the screened string tension,
though a slightly smaller value, say $T=(0.28-0.30) GeV^2$ is found.

These studies might indicate that the color screened
quasi-confinement potential with a large string tension,
say, $T=(0.26-0.32)GeV^2$ should be an interesting possibility.

The following observations might be in order.

(1). In order to get better results for higher excited $Q\bar Q$
states (e.g. $\psi(4160)$ and $\psi(4415)$), a screened confining
potential plus a Coulomb potential (with fixed or running $\alpha_s$)
seem to work well. Whereas
the unscreened linear potential give too large level spacings and
leptonic widths. While a large
$\alpha_{s}$ with a normal string tension (e.g. as shown in (7))
is possible, a smaller $\alpha_{s}$, which is more consistent with
the value of QCD scale parameter, with a larger string tension (e.g.
as shown in (8) and (9)) seem to work better for both $c\bar c$
and $b\bar b$ states. The screening parameter $\mu$ is found to
be $(0.14\pm 0.03)GeV$. This value is consistent with lattice QCD
calculations$^{[7]}$.

(2). In our calculation we have simply focused on the spin-independent
solutions of the Schrodinger eqution including the mass spectra
and the leptonic widths, and ignored the coupled channel effects
and relativistic corrections. In fact, the coupled channel effects
(see e.g. refs.[5,8]) and the
relativistic corrections (see e.g. refs.[15,19,20,21,22]) within
the linear
potential model seem to be unable to solve the puzzle regarding
$\psi(4160)$ and $\psi(4415)$, as well as some other highly excited
states. For instance, in the linear confinement model of ref.[21],
with relativistic corrections
the masses are found to be (in units of $MeV$) 3097, 3527, 3681, 3846,
4108, and 4446, for $1S, 1P, 2S, 1D, 3S$, and $4S$ $c\bar c$
states respectively.
We see that although relativistic corrections for
$c\bar c$ are important ( the energy shifts due to relativistic
corrections ranging from $-61MeV$ to
$-219MeV$ from $1S$ to $4S$ states), the energy spacings
with relativistic corrections can be very similar to that obtained
without relativistic corrections  e.g. in the Cornell
model$^{[8]}$ (the model in ref.[21] will of
course have different parameters from that in ref.[8]). This may imply
that as far as the energy spacings are concerned
the relativistic effects
may be largely absorbed by the readjustment of potential parameters
( e.g., the value of string tension takes $T=0.22GeV^2$ in ref.[21]
while $T=0.18GeV^2$ in ref.[8] ) and therefore
the relativistic effects appear to be small in practice.
Although this result is seen
specifically in the model of ref.[21], the conclusion here can be quite
general and similar observations have also been made by other
authors (see, e.g., ref.[22]). Hence the relativistic corrections
with unscreened confining potential are expected to be
not very helpful in solving
the difficulties associated with e.g. $\psi(4160)$ and $\psi(4415)$.
So it is very likely that in order to
improve the fit to the higher excited states the screened confining
potential is still needed even with these
coupled channel and relativistic effects taken into consideration.

(3). We have tried to calculate the spin-dependent splittings
of these heavy quarkonium states in a very simple version. If the
spin-spin force is entirely due to the lowest order perturbative
one-gluon exchange, the $0^{-}-1^{-}$ meson
mass splitting $\Delta$ will be given by
$$
\Delta=\frac{32\pi\alpha_s}{9m_{Q}^2}\left |\Psi(0)\right |^2.%
\eqno (12)
$$
Then for the $J/\psi$ and $\eta_{c}$, with
$\alpha_{s}=0.306, m_c=1.6GeV$ as given in (8),
and the Schrodinger wave
functions obtained by using (8), we get a mass
spliting $\Delta=110 MeV$,
slightly smaller than its experimental value $(118\pm 2)MeV$.
As for the fine splittings the situation is more complicated, since
the long-range nonperturbative forces may contribute.
It is argued$^{[23]}$ based on a consistent condition due to Lorentz
invariance that the confining potential should transform as a Lorentz
scalar and therefore induce a spin-orbit term which then compensates
the short-ranged spin-orbit force caused by one-gluon exchange.
On the other hand, in many studies regarding chiral symmetry
breaking, in order
to preserve chiral invariance the vector ( or at least the time component
of a 4-vector) confining force has to be chosen$^{[24]}$. In the
phenomenological studies of heavy quarkonium spectra (see, e.g.,
refs.[15,19,20,21,25,26]), though the scalar confining is favored a
vector-scalar mixture for the confining potential may work even better.
For instance, the vector-scalar mixed confining potential in some
models can give an excellent fit to the low lying $c\bar c$
and $b\bar b$
spectra$^{[25]}$. It is also argued that the observed tiny mass
difference ( about $-0.9MeV$ )$^{[4]}$ between the center of gravity
of triplet $1P$ and the singlet $1P$
charmonium states may not necessarily
mean the
short-ranged perturbative hyperfine splitting is dominant because
nonperturbative and other effects could be also important${[25,26]}$.
Although there are uncertainties for the spin-dependent splittings
and the Lorentz transformed structure of the confining potential
especially the color screened quasi-confining potential, we believe
it should be dominated by the scalar with possibly a small mixture
of the vector component.  We find for the P-wave mass splittings,
if the screened confining potential is a pure scalar,
then with (5), (8), and (9) the obtained splittings are too small.
If it is a vector-scalar
mixture with a weight factor being about 3:7 then a fairly good fit
can be obtained. However, we would like to emphasize that to calculate the
spin-dependent splittings the naive calculation given here with
a Coulomb potential of fixed $\alpha_s$ in (5) with (8) and (9)
should not be a good one, and a more refined calculation with higher
order perturbative corrections and nonperturbative effects is
apparently better. Therefore our calculation for the spin-dependent
splittings is not conclusive, and we will leave this to
a more refined work.

To conclude, by studying the heavy quarkonium spectra especially
for the
higher excited states, e.g. $\psi (4160), \psi (4415),$ and
$\Upsilon (11020)$, we find some evidence for the color screened
confining potential, which is expected theoretically when the
creation of dynamical
light quark pairs at large distances is taken into consideration.
A large string tension, say $T=(0.26-0.32) GeV^2$ is favored by an
overall fit to the mass spectra and leptonic widths. The existing
calculations with the coupled channel effects and
relativistic corrections based on the unscreened linear potential
model seem to unable to solve the difficulties associated with those
higher excited states. Therefore the color screened linear
confining potential with a large string tension should be an
interesting possibility and deserve further investigations.

This work was supported in part by the National Natural Science
Foundation of
China and the State Education Commission of China.

\newpage

%\begin{center}

Table 1: Calculated masses and leptonic widths for
charmonium states
with the screened potential (5) and parameters (8), where $\Gamma_{ee}
=\Gamma^0_{ee}(1-\frac{16}{3 \pi}\alpha_s(m_c))$ with
$\alpha_s(m_c)=0.28^{[16]}$.

%\end{center}

\vspace{0.3cm}
%\begin{eqnarray*}
\begin{tabular}{|l|l|l|l|l|l|}
\hline
States & Mass(MeV) & $\Gamma^0_{ee}$(keV) &
$\Gamma_{ee}$(keV) &
$\Gamma^{exp.}_{ee}$(keV) &  Candidate  \\
\hline
1S &  3097 &  10.18 & 5.34 & $5.26\pm 0.37$ & $\psi$(3097) \\
\hline
2S & 3686 & 4.13  &2.17 & $2.14\pm 0.21$ &  $\psi$(3686) \\
\hline
3S & 4033 & 2.35  & 1.23 & $0.75\pm 0.15$ & $\psi$(4040) \\
\hline
4S &  4262 &   1.46  & 0.77  & $0.77\pm 0.23$ & $\psi$(4160) \\
\hline
5S & 4415  &  0.91 & 0.48 & $0.47\pm 0.10$  &
$\psi$(4415) \\
\hline
1P &3526 & \ &\ &\ &   $\chi(3526)_{c.o.g}$\\
\hline
1D &3805  &\ &\ &\  &    $\psi(3770)$\\
\hline
2D &4105  &\ &\ &\ &\\
\hline
\end{tabular}
%\end{eqnarray*}
\vspace{1cm}
%\begin{center}

 Table 2: Calculated masses and leptonic widths for bottomonium states
with the screened potential (5) and parameters (9), where
$ \Gamma_{ee}=
\Gamma^0_{ee}(1-\frac{16}{3 \pi}\alpha_s(m_b)) $ with
$\alpha_s(m_b)=0.19^{[16]}$.

%\end{center}

\vspace{0.3cm}

\begin{tabular}{|l|l|l|l|l|l|}
\hline
States &Mass(MeV)&
$\Gamma^0_{ee}$(keV) & $\Gamma_{ee}$(keV)&
$\Gamma^{exp.}_{ee}$(keV) &  Candidate \\
\hline

1S &  9460 &  1.94 & 1.31 &  $1.32\pm 0.03$ &  $\Upsilon$ (9460) \\
\hline
2S & 10023 &  0.90 & 0.61 &  $0.58\pm 0.10$ &  $\Upsilon$ (10023)
\\
\hline

3S & 10368 &  0.62 & 0.42 &  $0.47\pm 0.06$ &  $\Upsilon$ (10355)
 \\
\hline
 4S & 10627 &  0.47 & 0.32 & $ 0.24\pm 0.05$ &  $\Upsilon$ (10580)
 \\
\hline
5S & 10833 &  0.37 & 0.25 &  $0.31\pm 0.07$ &  $\Upsilon$ (10860)
 \\
\hline
6S & 11002 &  0.30 & 0.20 &  $0.13\pm 0.03$ &  $\Upsilon$ (11020)
\\
\hline
1P &   9894 &       &      &      & $\chi_{b}(9900)_{c.o.g.}$
\\
\hline
2P &  10267 &       &      &      & $\chi_{b}(10261)_{c.o.g.}$
\\
\hline
1D &  10152 &       &      &       &         \\
\hline
2D&  10451 &       &      &        &    \\
\hline
\end{tabular}
%\end{eqnarray*}
\end{document}